\documentclass[preprint]{elsarticle}
\usepackage{amssymb}
\usepackage{amsmath}
\usepackage{xcolor}
\RequirePackage{graphicx}
\usepackage{adjustbox}
\journal{Nuclear Physics B}

\begin{document}

\begin{frontmatter}

%% Title, authors and addresses

%% use the tnoteref command within \title for footnotes;
%% use the tnotetext command for theassociated footnote;
%% use the fnref command within \author or \address for footnotes;
%% use the fntext command for theassociated footnote;
%% use the corref command within \author for corresponding author footnotes;
%% use the cortext command for theassociated footnote;
%% use the ead command for the email address,
%% and the form \ead[url] for the home page:
%% \title{Title\tnoteref{label1}}
%% \tnotetext[label1]{}
%% \author{Name\corref{cor1}\fnref{label2}}
%% \ead{email address}
%% \ead[url]{home page}
%% \fntext[label2]{}
%% \cortext[cor1]{}
%% \affiliation{organization={},
%%             addressline={},
%%             city={},
%%             postcode={},
%%             state={},
%%             country={}}
%% \fntext[label3]{}

\title{Generalized Half-Dyon in Weinberg-Salam Theory}

%% use optional labels to link authors explicitly to addresses:
%% \author[label1,label2]{}
%% \affiliation[label1]{organization={},
%%             addressline={},
%%             city={},
%%             postcode={},
%%             state={},
%%             country={}}
%%
%% \affiliation[label2]{organization={},
%%             addressline={},
%%             city={},
%%             postcode={},
%%             state={},
%%             country={}}

\author[1]{Guo-Quan Wong }

\author[1]{Khai-Ming Wong\corref{cor1}%
		}
\ead{kmwong@usm.my}        

\author[1]{Dan Zhu}
       
\cortext[cor1]{Corresponding author}

\affiliation[1]{organization={School of Physics, Universiti Sains Malaysia},
           % addressline={},%Department and Organization}, 
            %city={},
            postcode={11800 USM}, 
            state={Penang},
            country={Malaysia}}

\begin{abstract}
%% Text of abstract
We construct and study numerical solutions corresponding to generalized electrically charged half-monopole in Weinberg-Salam theory, denoted as Type I and Type II solutions. These solutions possess magnetic charge $q_m = +2 n \pi/e$ ($-2 n \pi/e$) and electric charge $q_{e}$ that depends on the electric charge parameter $\eta$, as well as net zero neutral charge. Other properties of this half-dyon configurations such as magnetic dipole moment and angular moment are studied. The energy of this half-dyon configuration is infinite due to singularity at the location of the half-dyon.

\end{abstract}

%%Graphical abstract
%\begin{graphicalabstract}
%\includegraphics{grabs}
%\end{graphicalabstract}

%%Research highlights
%\begin{highlights}
%\item Research highlight 1
%\item Research highlight 2
%\end{highlights}

\begin{keyword}
%% keywords here, in the form: keyword \sep keyword
Weinberg-Salam theory \sep Cho-Maison monopole \sep half-dyon
%% PACS codes here, in the form: \PACS code \sep code

%% MSC codes here, in the form: \MSC code \sep code
%% or \MSC[2008] code \sep code (2000 is the default)

\end{keyword}

\end{frontmatter}

%% \linenumbers

%% main text
\section{Introduction}
\label{sec:1}

Magnetic monopole has always been a subject that attracts a lot of interest. P.A.M. Dirac is the first to introduce magnetic monopole in Maxwell's theory \cite{kn:1}. Wu and Yang \cite{kn:2} then generalized the idea of magnetic monopole to non-Abelian gauge theories. However, both Dirac and Wu-Yang monopole possess infinite energy as they contain singularity at the origin. The first breakthrough came when 't Hooft and Polyakov independently discovered a regular monopole solution in SU(2) Yang-Mills-Higgs (YMH) theory \cite{kn:3}. The 't Hooft-Polyakov monopole possesses finite energy as the gauge potential is well-defined in all space. The mass of 't Hooft-Polyakov monopole was estimated to be of order 137 $M_{\scalebox{.5}{\mbox{W}}}$, where $M_{\scalebox{.5}{\mbox{W}}}$ is the mass of intermediate vector boson.

Besides the spherically symmetric 't Hooft-Polyakov monopole, there exists other interesting axially symmetric monopole configurations in SU(2) YMH theory. These include the single $n$-monopole \cite{kn:4}, system of monopole-antimonopole pair (MAP), monopole antimonopole chain (MAC), and vortex-ring configurations \cite{kn:5}. The solutions in Ref. \cite{kn:5} possess finite energy and represent a chain of magnetic monopoles lying in alternating order along the symmetrical axis. Recently axially symmetric finite energy half-monopole configurations are also found to exist in the SU(2) YMH theory \cite{kn:6}. Later it is found that this half-monopole can coexist with another ’t Hooft–Polyakov monopole \cite{kn:7}. 

We recently reported a generalized half-dyon solution in SU(2) YMH theory in Ref.~\cite{su2-half} by extending the work from Ref.~\cite{kn:6}. In this paper, we considered the half-dyon with varying $\phi$-winding number $n$, electric charge parameter $\eta$, and Higgs self-coupling constant $\beta$. There are two types of solutions, denoted as Type I and Type II in which the Type I solution refers to half-dyon configuration with positive magnetic charge equal to $2n\pi/g$, where $g$ is the coupling constant, that extends along the negative $z$-axis. On the contrary, Type II solution is a half-dyon configuration with negative magnetic charge ($-2n\pi/g$) positioned along the positive $z$-axis. Both Type I and Type II half-dyon configurations appear to be the reflection of each other about the $\rho = \sqrt{x^2+y^2}$ plane. The fundamental properties of both configurations such as total energy, electric charge and magnetic dipole moment are addressed.

The SU(2) $\times$ U(1) Weinberg-Salam theory possesses topological magnetic monopole solution, which is simply known as the electroweak monopole or Cho-Maison monopole \cite{kn:8}. The Cho-Maison monopole describes a real monopole dressed by the physical W-boson and Higgs field, and can be viewed as a hybrid between Dirac monopole and 't Hooft-Polyakov monopole. Although the Cho-Maison monopole has infinite energy due to the U(1) singularity at the origin,  the mass of this monopole can be regularized and estimated at the range of 4 to 10 TeV \cite{kn:9,kn:10,kn:11}. It is also recently shown that there is a more natural way to regularize the energy, suggesting that the new BPS bound for the Cho-Maison monopole may not be smaller than 2.98 TeV, more probably 3.75 TeV \cite{kn:12}. As stated also in Ref. \cite{kn:12}, the discovery of electroweak monopole serves as an important topological test of the standard model. This makes the experimental detection of electroweak monopole an urgent issue after the discovery of Higgs boson, with experimental detectors actively searching for magnetic monopole \cite{kn:13,kn:14,kn:15,kn:16}. 

Another line of research in SU(2) $\times$ U(1) Weinberg-Salam theory includes the work of Y. Nambu \cite{kn:17} that demonstrates the existence of a pair of monopole and antimonopole bounded by flux string of the $Z_0$ field. The total energy of this MAP configuration is finite and the mass of the system (monopole, antimonopole together with the string) is estimated to be in TeV range. The real electromagnetic field is a linear combination of U(1) and SU(2) gauge fields at asymptotic large distance. Although the arguments and calculations were not rigorous, the existence of such string-like configurations attracts a great deal of interest. 

There are also huge amount of work done on the classical solutions of SU(2) $\times$ U(1) Weinberg-Salam theory. These are the `sphaleron' which possesses baryon number $Q_{\scalebox{.6}{\mbox{B}}} = 1/2$, and it is also found that there is an electric current in the U(1) field \cite{kn:18,kn:19}. On the other hand, the works of Hindmarsh and James \cite{kn:20}, and Radu and Volkov \cite{kn:21} also found that within the sphaleron there is a monopole-antimonopole pair and loop of electromagnetic current. Other sphaleron configurations of the Weinberg–Salam theory include the system of sphalerons lying along symmetrical axis \cite{kn:22,kn:23}, which possess magnetic dipole moment and finite energy. Though possessing axial symmetry, the inner structure of sphaleron is not really revealed. 

\sloppy{
Recently more magnetic monopole solutions were found in the SU(2) $\times$ U(1) Weinberg–Salam theory by using a generalized axially symmetric ansatz of that in Ref. \cite{kn:8}. These are the MAP, MAC, and vortex-ring configurations with axial symmetry \cite{kn:24}. It was shown explicitly that the MAP/vortex-ring configurations which possess zero net magnetic charge are actually sphaleron (M-A) and sphaleron–antisphaleron pair (M-A-M-A), hence confirming that the sphaleron found by others \cite{kn:18,kn:19,kn:20,kn:21,kn:22,kn:23} does possess inner structure. The monopole and antimonopole in the sphaleron possess magnetic charges $ \pm 4 \pi \sin^2\theta_{\scalebox{.6}{\mbox{W}}} / e$ respectively and they are half Cho–Maison monopole (antimonopole) if one considers Weinberg angle $\theta_{\scalebox{.6}{\mbox{W}}} = \pi/4$. The MAC/vortex-ring configurations that possess net magnetic charge $4 \pi/e$ is a sequence of Cho-Maison monopole (with antimonopole) chain. The single Cho–Maison monopole \cite{kn:8} is the first member of this sequence of solutions.}

Additionally, axially symmetrical half-monopole configuration of the Weinberg–Salam theory are also reported \cite{kn:25}. The equations of motion are solved numerically for all space when the $\phi$-winding number $n = 1$. The solution is studied by varying the Weinberg angle $\theta_{\scalebox{.6}{\mbox{W}}}$ from $\pi/18$ to $\pi/2$, when the Higgs field self-coupling constant is $\lambda = 1$, and also by varying the Higgs field self-coupling constant $\lambda$ when the Weinberg angle $\sin^2\theta_{\scalebox{.6}{\mbox{W}}} = 0.2312$. The Higgs vacuum expectation value and the unit electric charge $e$ are both set to unity. This half-monopole solution possesses magnetic charge $2 \pi/e$ and hence could be considered a half Cho–Maison monopole. It possesses finite total energy even though the electromagnetic gauge potential is singular along the negative $z$-axis. In terms of structure, this half-monopole is a finite length line magnetic charge extending from the origin $ r = 0 $ and lying along the negative $z$-axis. There are however some drawbacks in \cite{kn:24} and \cite{kn:25}, i.e. dimension is not considered (the gauge coupling constants are set to unity) and unphysical Weinberg angle are used for ease of calculation \cite{kn:24}. These drawbacks hinder important physical information to be extracted. To address the issues, correct dimension is considered in this work.

In this paper, we extend the work in Refs.~\cite{su2-half} and \cite{kn:25} by considering inclusion of electric charge to the half-monopole in SU(2) $\times$ U(1) Weinberg-Salam theory, creating the so-called half-dyon solutions. Following the definition for Type I and Type II half-dyon solutions in Ref.~\cite{su2-half}, we construct similar Type I and Type II half-dyon solutions in SU(2) $\times$ U(1) Weinberg-Salam theory. Here we consider $\phi$-winding number $1 \leq n \leq 3$, physical Weinberg angle $\sin^2\theta_{\scalebox{.6}{\mbox{W}}} = 0.2229$, Higgs self-coupling constant $\beta = 0.77818833$, and electric charge parameter $\eta$ valued from zero up to a critical $\eta_c$. Numerical solutions exist only when $0 \leq \eta \leq \eta_{c}$ and they cease to exist when $\eta > \eta_{c}$.  The Type I half-dyon solutions lie along the positive $z$-axis with total magnetic charge $+2n\pi/e$ and electric charge $q_e$ that depends on $\eta$. The Type II half-dyon solutions, on the other hand, has total magnetic charge $-2n\pi/e$ and lie along the positive $z$-axis. The electromagnetic gauge potential of the system is singular along the negative $z$-axis . Moreover, the energy of the U(1) part is singular at the location of the half-dyon hence the energy of the configuration is infinite.

The paper is organized as follows. In Section 2 we briefly discuss the Weinberg-Salam theory. The numerical method used to acquire the solutions will be discussed in Section 3, which includes the axially symmetric ansatz used to obtain the reduced equations of motion. In Section 4, we present some formulations of the solutions' fundamental properties, i.e. total energy, electromagnetic, neutral fields etc. The half-dyon solutions are analyzed and discussed in Section 5. We end with some comments in Section 6.

\section{The Standard Weinberg-Salam Theory}
\label{sec:2}

We consider the SU(2) $\times$ U(1) Weinberg-Salam Lagrangian as
\begin{equation}
{\cal L}_{M} = -\frac{1}{4}F^a_{\mu\nu} F^{a\mu\nu} - \frac{1}{4} G_{\mu\nu} G^{\mu\nu}  - \left( {\cal D}_\mu \phi \right)^{\dagger} \, {\cal D}^\mu \phi - V \left( \phi \right),
\label{eq.1}
\end{equation}
where
\begin{equation}
V\left(\phi \right) = - \frac{\lambda}{2} \left( \phi^{\dagger} \phi- \frac{\mu^2}{\lambda} \right)^2,
\label{eq.2}
\end{equation}
and
\begin{align}
{\cal D}_{\mu} \phi &= \left( \partial_{\mu} - \frac{ig}{2} \sigma^a A^a_{\mu} - \frac{ig'}{2} B_{\mu} \right)  \phi \nonumber\\ 
&= \left( D_{\mu} - \frac{ig'}{2} B_{\mu}  \right) \phi,
\label{eq.3}
\end{align}
in which ${\cal D}_{\mu}$ is the covariant derivative of the SU(2) $\times$ U(1) group and $D_{\mu}$ is the covariant derivative of the SU(2) group only. The SU(2) gauge coupling constants, potentials and field tensors are given by $g$, $A^{a}_{\mu}$ and $F^{a}_{\mu\nu}$, whereas those of the U(1) group are given by $g'$, $B_{\mu}$ and $G_{\mu\nu}$. The term $\sigma^a$ is the Pauli matrices,
\begin{equation}
\sigma^a = \begin{bmatrix}
    \delta^{a}_{3} & \delta^{a}_{1} - i \delta^{a}_{2} \\
    \delta^{a}_{1} + i \delta^{a}_{2} & -\delta^{a}_{3}
  \end{bmatrix},   
\label{eq.4}
\end{equation} %{\scalebox{.5}{\mbox{H}}}
whereas $\phi$ is the complex Higgs doublet and $\lambda$ is the Higgs self-coupling constant. The mass of the Higgs boson is given by $m_\text{{H}} = \sqrt{2} \mu $, where $H_0 = \sqrt{2} \mu / \sqrt{\lambda}$ is the Higgs vacuum expectation value. 

From Lagrangian (\ref{eq.1}), the equations of motion are
\begin{align}
{\cal D}^{\mu} {\cal D}_{\mu} \phi &= \lambda \left( \phi^{\dagger}\phi - \frac{\mu^2}{\lambda} \right) \phi, \nonumber\\
 D^{\mu} F^{a}_{\mu\nu} &= \frac{ig}{2} \left[ \phi^{\dagger} \sigma^a \left( {\cal D}_{\nu} \phi \right) - \left( {\cal D}_{\nu} \phi \right)^{\dagger} \sigma^a \phi  \right], \nonumber\\
\partial^{\mu} G_{\mu\nu} &= \frac{i g'}{2} \left[ \phi^{\dagger} \left( {\cal D}_{\nu} \phi \right) -  \left( {\cal D}_{\nu} \phi \right)^{\dagger} \phi  \right].
\label{eq.5}
\end{align}
The Higgs doublet is defined as
\begin{align}
\phi = \frac{{\cal H}}{\sqrt{2}} ~\xi,~~  \left(  \phi^{\dagger} \phi = \frac{{\cal H}^2}{2}, ~~\xi^{\dagger} \xi = 1 \right).
\label{eq.6}
\end{align}
We also define the rectangular coordinate system unit vectors as
\begin{align}
\hat{n}^a &= h_1 \, \hat{r}^a + h_2 \, \hat{\theta}^a \nonumber\\
&= \cos\left( \alpha - \theta  \right) \, \hat{r}^a + \sin\left( \alpha - \theta  \right) \nonumber\\
&= \sin \alpha \cos n\phi \, \delta^{a1} + \sin \alpha \sin n\phi \, \delta^{a1} + \cos\alpha \, \delta^{a3}.
\label{eq.7}
\end{align}
The functions $\cos\alpha$ and $\sin\alpha$ are defined as
\begin{align}
&\cos\alpha = \frac{\Phi_1 \cos\theta - \Phi_2 \sin\theta}{\sqrt{\Phi_1^2+\Phi_2^2}} = h_1 \cos\theta - h_2 \sin\theta, \nonumber\\ 
&\sin\alpha = \frac{\Phi_1 \sin\theta + \Phi_2 \cos\theta}{\sqrt{\Phi_1^2+\Phi_2^2}} = h_1 \sin\theta + h_2 \cos\theta.
\label{eq.8}
\end{align}
The Higgs field is
\begin{equation}
\Phi^a = \Phi_1 \, \hat{r}^a + \Phi_2 \, \hat{\theta}^a = {\cal H} \, \hat{\Phi}^a,
\label{eq.9}
\end{equation}
where ${\cal H} =  \sqrt{\Phi^2_1 + \Phi_2^2}$ and the Higgs unit vector can be written as
\begin{equation}
\hat{\Phi}^a = - \xi^{\dagger} \sigma^a \xi = \hat{n}^a.
\label{eq.10}
\end{equation}

Our previous works in the SU(2) YMH theory reveals that the angle $\alpha \left( r,\theta \right) \rightarrow p \theta$ as $r \rightarrow \infty$, where $p = 1, 2, 3, ....$ is a natural number representing the number of magnetic poles (monopoles and antimonopoles) in the configurations. When $p$ is odd, one obtains the MAC solutions and when $p$ is even, we obtain the MAP solution. Monopole solution with half-integer charge arises when $p$ is half-integer, i.e. when $p = 1/2$, we get the half-monopole solution and when $p = 3/2$, we obtain the one plus half-monopole solution. Similar phenomena is applicable to the case of Weinberg-Salam theory.

\section{The Numerical Method}
\label{sec:3}

In this paper we consider the following electrically charged axially symmetric ansatz,
\begin{align}
& \phi = \frac{{\cal H}}{\sqrt{2}} ~\xi,~~~\xi = i \begin{bmatrix}
    \sin\frac{\alpha}{2} \, e^{- i n  \phi} \\
    -\cos\frac{\alpha}{2} 
  \end{bmatrix},   \nonumber\\
& g A^a_0 = \tau_1 \, \hat{r}^a + \tau_2 \, \hat{\theta}^a, \nonumber\\
& g A^a_{i} = - \frac{\psi_1}{r} \, \hat{\phi}^a \hat{\theta}_i + \frac{n P_1 }{r \sin\theta} \, \hat{\theta}^a \hat{\phi}_i + \frac{R_1}{r} \, \hat{\phi}^a \hat{r}_i -  \frac{n P_2}{r \sin\theta} \, \hat{r}^a \hat{\phi}_i,  \nonumber\\
& g' B_0 = \tilde{B}_0,~~ g' B_i = \frac{n B_s}{r \sin\theta} \, \hat{\phi}_i. 
\label{eq.11}
\end{align}
Here $\psi_1$, $P_1$, $R_1$, $P_2$, $B_s$, ${\cal H}$, $\tilde{B}_0$, $\tau_1$ and $\tau_2$ are all functions of $r$ and $\theta$. The spatial spherical coordinate unit vectors are
\begin{align}
\hat{r}_{i} &= \sin\theta \, \cos\phi \, \delta_{i1} + \sin\theta \, \sin\phi \, \delta_{i2} + \cos\theta \, \delta_{i3}, \nonumber\\
\hat{\theta}_{i} &= \cos\theta \, \cos\phi \, \delta_{i1} + \cos\theta \, \sin\phi \,\delta_{i2} - \sin\theta \, \delta_{i3}, \nonumber\\
\hat{\phi}_{i} &= - \sin\phi \, \delta_{i1} + \cos\phi \, \delta_{i2},
\label{eq.12}
\end{align}
whereas the isospin coordinate unit vectors are 
\begin{align}
\hat{r}^{a} &= \sin\theta \, \cos n\phi \, \delta^{a}_{1} + \sin\theta \, \sin n\phi \, \delta^{a}_{2} + \cos\theta \, \delta^{a}_{3}, \nonumber\\
\hat{\theta}^{a} &= \cos\theta \, \cos n\phi \, \delta^{a}_{1} + \cos\theta \, \sin n\phi \, \delta^{a}_{2} - \sin\theta \, \delta^{a}_{3}, \nonumber\\
\hat{\phi}^{a} &= - \sin n\phi \, \delta^{a}_{1} + \cos n\phi \, \delta^{a}_{2}.
\label{eq.13}
\end{align}

Following Eq.(\ref{eq.5}), the equations for Higgs field are
\begin{align}
& \partial^{i} \partial_{i} {\cal H} - \frac{\lambda}{2} \left( {\cal H}^2 - \frac{2 \mu^2}{\lambda} \right) {\cal H} \nonumber\\
& - \frac{{\cal H}}{4}  \left[  \left(  -\frac{\psi_1}{r} + \frac{\dot{\alpha}}{r} \right)^2 + \left( \frac{R_1}{r} - \alpha' \right)^2  \right] \nonumber\\
& + \frac{{\cal H}}{4 r^2} \left[ \left( \frac{A_s + n}{\sin\theta} \right)^2 - \left( \frac{A_s - n B_s}{\sin\theta} \right)^2  \right] \nonumber\\
& + \frac{{\cal H}}{4 r^2} \left[ - \left( \frac{n P_1}{\sin\theta} - n \right)^2 - \left( \frac{n P_2}{\sin\theta} - n \cot\theta\right)^2 \right] \nonumber\\
& + \frac{{\cal H}}{4} \left[ \left( \tau_1 h_2 - \tau_2 h_1  \right)^2 + \left( \tilde{B}_0 - \tau_1 h_1 - \tau_2 h_2  \right)^2 \right] = 0
\label{eq.14}
\end{align}
and 
\begin{align}
& \frac{\cot\theta}{r^2} -\left( \frac{\partial^i \partial_i h_1}{h_2} - \frac{ \partial^i h_1 \partial_i h_2}{h^2_2} \right) \nonumber\\
& - \frac{1}{r^2} \left( \dot{ \psi_1} + \psi_1 \cot\theta \right) + \frac{1}{r^2} \left( r R_1' + R_1 \right) \nonumber\\
& + \frac{2}{r} \left( R_1 - r \frac{h_1'}{h_2}  \right) \frac{{\cal H}'}{{\cal H}} - \frac{2}{r^2} \left[ \psi_1 - \left( 1 - \frac{\dot{h_1}}{h_2}  \right) \right] \frac{\dot{{\cal H}}}{{\cal H}} \nonumber\\
& + \frac{n^2 \left( B_s + 1 \right)}{r \sin\theta} \left( \frac{P_1 h_1 + P_2 h_2}{\sin\theta} - h_1 - h_2 \cot\theta \right) \nonumber\\
& + \tilde{B}_0 \left( \tau_1 h_2 - \tau_2 h_1 \right) = 0.
\label{eq.15}
\end{align}
The equations of motion for SU(2) gauge field are
\begin{align}
& \frac{1}{g} D^{i} F^{a}_{ij} = \frac{1}{g} \left( \partial^{i} F^{a}_{ij} + \epsilon^{abc} g A^{b i } F^{c}_{ij} + \epsilon^{abc} g A^{b 0 } F^{c}_{0j} \right)  \nonumber\\
& = \frac{ {\cal H}^2}{4r} \left[ \frac{ \left( A_s - n B_s \right)}{\sin\theta} \hat{n}^a \hat{\phi}_{j}  \right] \nonumber\\
& + \frac{ {\cal H}^2}{4r} \left[ n \left( \frac{  P_1 h_1 + P_2 h_2 }{\sin\theta} -  h_1 - h_2 \cot\theta \right) \hat{n}^{a}_{\perp} \hat{\phi}_{j}   \right] \nonumber\\
& + \frac{ {\cal H}^2}{4r} \left[ \left( R_1 + r \alpha'  \right) \hat{\phi}^a \hat{r}_j - \left[ \psi_1 - \dot{\alpha}  \right]   \hat{\phi}^a \hat{\theta}_j  \right],
\label{eq.16}
\end{align}
and
\begin{align}
& \frac{1}{g} D^{i} F^{a}_{i0} = \frac{1}{g} \left( \partial^{i} F^{a}_{i0} + \epsilon^{abc} g A^{b i } F^{c}_{i0} \right) \nonumber\\
& = \frac{{\cal H}^2 }{4r}\left[ \left( \tau_1 h_1 + \tau_2 h_2 - \tilde{B}_0 \right) \hat{n}^{a} + \left( \tau_2 h_1 - \tau_1 h_2  \right) \hat{n}^{a}_{\perp}   \right], \nonumber\\
\label{eq.17}
\end{align}
where 
\begin{equation}
\alpha' = - \frac{h_1'}{h_2}, ~ \dot{\alpha} = 1 - \frac{\dot{h_1}}{h_2},
\label{eq.18}
\end{equation}
and the unit vector $\hat{n}^{a}_{\perp} = -h_2 \, \hat{r}^{a} + h_1 \, \hat{\theta}^{a}$ is perpendicular to $\hat{n}^{a}$. The equations for U(1) gauge field are
\begin{equation}
\partial^{i} \partial_{i} \left( \frac{n B_s}{r \sin\theta} \right) - \frac{n}{r^3 \sin^3\theta} B_s = \frac{g'^2}{4r} {\cal H}^2 \frac{ \left( n B_s - A_s \right)}{\sin\theta},
\label{eq.19}
\end{equation}
and
\begin{equation}
\partial^{i} \partial_{i} \tilde{B}_0 = \frac{g'^2}{4} {\cal H}^2 \left( \tilde{B}_0 - \tau_1 h_1 - \tau_2 h_2  \right).
\label{eq.20}
\end{equation}
Here prime denotes derivative with respect to $r$, dot denotes derivative with respect to $\theta$ and also $H_0 = \sqrt{2} \mu/\sqrt{\lambda}$ is the Higgs vaccum expectation value. The function $A_s$ is given by
\begin{equation}
A_s = n \left\{ P_1 h_2 -  P_2 h_1 -  \left( 1 - \cos\alpha \right) \right\}.
\label{eq.21}
\end{equation}
There are all together ten equations of motion in (\ref{eq.14})-(\ref{eq.17}), which will be solved numerically. To facilitate numerical calculation, we consider the dimensionless coordinate $x = M_{\scalebox{.5}{\mbox{W}}} r$, where $M_{\scalebox{.5}{\mbox{W}}} =  g H_0/2$ and the following transformed functions
\begin{align}
& H \rightarrow H_0 H, ~\tilde{B}_0 \rightarrow g' H_0 \tilde{B}_0, \nonumber\\
& \tau_1 \rightarrow g H_0 \tau_1, ~ \tau_2 \rightarrow g H_0 \tau_2.
\label{eq.22}
\end{align}
The equations of motion then depend on coupling constant $\beta$ and Weinberg angle $\theta_{\scalebox{.5}{\mbox{W}}}$, where
\begin{equation}
\beta^2 = \frac{\lambda}{g^2},~~ \omega = \tan\theta_{\scalebox{.5}{\mbox{W}}} = \frac{g'}{g}.
\label{eq.23}
\end{equation} 
Here we consider physical value of $\omega = 0.53557042$ by adopting $\sin^2\theta_{W} = 0.2229$. Since $M_{\scalebox{.5}{\mbox{H}}} = \sqrt{2} \mu$ and $M_{\scalebox{.5}{\mbox{W}}} = \frac{1}{2}g H_0$, we may put 
\begin{equation}
\beta = \frac{1}{2} \frac{M_{\scalebox{.5}{\mbox{H}}}}{M_{\scalebox{.5}{\mbox{W}}}}, 
\label{eq.24}
\end{equation} 
where by adopting $M_{\scalebox{.5}{\mbox{H}}} = 125.10$ GeV and $M_{\scalebox{.5}{\mbox{W}}} = 80.379$ GeV, the physical value of $\beta$ used here is 0.77818833.

The half-dyon solutions are solved numerically by considering physical Weinberg angle $\sin^2\theta_{\scalebox{.5}{\mbox{W}}} = 0.2229$ and Higgs self coupling constant $\beta = 0.77818833 $, $\phi$-winding number $n$ ranging from $n=1$ to $n=3$, and electric charge parameter $0 \leq \eta \leq \eta_{c}$. The value of $\eta_{c}$ differs for each case of $n$, where $\eta_{c, n = 1} = 0.5001$, $\eta_{c, n = 2} = 0.5005$, and $\eta_{c, n = 3} = 0.1726$. 

The numerical procedures are also subject to the following boundary condition. At $r \rightarrow \infty $, the boundary condition is
\begin{align}
& \psi_1 = \frac{1}{2}, ~~P_1 = \sin\theta - \frac{1}{2} \left( \sin \frac{\theta}{2} \right) \left( 1 + \cos\theta \right), \nonumber\\
& R_1 = 0, ~~ P_2 = \cos\theta - \frac{1}{2} \left( \cos \frac{\theta}{2} \right) \left( 1 + \cos\theta \right), \nonumber\\
& \Phi_1 = \cos \frac{\theta}{2}, ~~\Phi_2 = - \sin \frac{\theta}{2}, \nonumber\\
& \tau_1 = \eta \cos \frac{\theta}{2}, ~~\tau_2 = - \eta \sin \frac{\theta}{2}, \nonumber\\
& \tilde{B}_0 = \frac{\eta}{\omega}, ~~n B_s = A_s = - \frac{n}{2} \left( 1 - \cos\theta \right).
\label{eq.25}
\end{align} 
Eq.(\ref{eq.25}) shows that at asymptotic large $r$ region, the time component of the gauge field and Higgs field are parallel in the isospin space. The asymptotic condition at small $r$ is the trivial vaccum solution 
\begin{align}
& \psi_1\left( 0,\theta \right) = P_1\left( 0,\theta \right) = R_1\left( 0,\theta \right)  = P_2\left( 0,\theta \right)  =0, \nonumber\\
& B_s \left( 0,\theta \right) = 0,~\partial_r \tilde{B}_0 \left( 0,\theta \right) = 0, \nonumber\\
& \sin\theta \, \Phi_1\left( 0,\theta \right) - \cos\theta \, \Phi_2\left( 0,\theta \right) = 0, \nonumber\\
& \sin\theta \, \tau_1\left( 0,\theta \right) - \cos\theta \, \tau_2\left( 0,\theta \right) = 0, \nonumber\\
& \left.  \frac{\partial }{\partial r }  \left\{ \cos\theta \, \Phi_1\left( r,\theta \right) - \sin\theta \, \Phi_2\left( r,\theta \right) \right\} \right|_{r = 0} = 0, \nonumber\\
& \left.  \frac{\partial }{\partial r }  \left\{ \cos\theta \, \tau_1\left( r,\theta \right) - \sin\theta \, \tau_2\left( r,\theta \right) \right\} \right|_{r = 0} = 0.
\label{eq.26}
\end{align} 
The boundary condition along the positive $z$-axis at $\theta = 0$ is
\begin{align}
& \partial_{\theta} \psi_1 = R_1 = P_1 = P_2 = \partial_{\theta} \Phi_1 = \Phi_2 = 0, \nonumber\\
& \partial_{\theta} \tau_1 = \tau_2 = B_s = \partial_{\theta} \tilde{B}_0 = 0,
\label{eq.27}
\end{align} 
whereas along the negative $z$-axis at $\theta = \pi$ it  is 
\begin{align}
& \partial_{\theta} \psi_1 = R_1 = P_1 = \partial_{\theta} P_2 = \Phi_1 = \partial_{\theta} \Phi_2 = 0, \nonumber\\
& \tau_1 = \partial_{\theta} \tau_2 = \partial_{\theta} B_s = \partial_{\theta} \tilde{B}_0 = 0.
\label{eq.28}
\end{align}

To obtain Type II solution, the boundary condition considered at $r \rightarrow \infty$ is
\begin{align}\label{boundary-cond-negative_1}
&\psi_1 = \frac{1}{2},~~ P_1 = \sin\theta - \frac{1}{2}\left(\cos\frac{\theta}{2}\right)(1-\cos\theta),\nonumber\\
&R_1 = 0,~~ P_2 = \cos\theta + \frac{1}{2}\left(\sin\frac{\theta}{2}\right)(1-\cos\theta),\nonumber\\
&\Phi_1 = -\sin\frac{\theta}{2},~~ \Phi_2 =-\cos\frac{\theta}{2},\nonumber\\
&\tau_1 = -\eta\sin\frac{\theta}{2},~~ \tau_2 =-\eta\cos\frac{\theta}{2}\nonumber\\
&\tilde{B}_0 = \frac{\eta}{\omega},~~ n B_s = A_s = - \frac{n}{2} \left( 1 + \cos\theta \right).
\end{align}
The same boundary condition, Eq.~(\ref{eq.26}), is used at $r = 0$. Along the positive $z$-axis ($\theta=0$), the boundary condition is 
\begin{align}\label{boundary-cond-negative_2}
&\partial_\theta\psi_1 = P_1 = R_1 = \partial_\theta P_2 = \Phi_1 = \partial_\theta\Phi_2 = 0 \nonumber \\
& \tau_1 = \partial_\theta\tau_2 = \partial_{\theta} B_s = \partial_{\theta} \tilde{B}_0 = 0,
\end{align}
and along the negative $z$-axis ($\theta=\pi$), the boundary condition is
\begin{align}\label{boundary-cond-negative_3}
&\partial_\theta\psi_1 = P_1 = R_1 = P_2 = \partial_\theta\Phi_1 = \Phi_2 = 0, \nonumber\\
& \partial_\theta\tau_1 = \tau_2 = B_s = \partial_{\theta} \tilde{B}_0 = 0.
\end{align}

The numerical calculations are mainly carried out using Maple and MATLAB software. The ten reduced second order partial differential equations (\ref{eq.14})-(\ref{eq.17}) are first transformed into a system of nonlinear equations by using finite difference approximation, based on the discretization on a non-equidistant grid of size $70 \times 60$ covering the integration regions $0 \leq \tilde{x} \leq 1$ and $0 \leq \theta \leq \pi$. Here $\tilde{x} = x/(x+1)$ is the compactified finite interval coordinate. The partial derivatives with respect to $x$ are then replaced accordingly, i.e. $\partial_x \rightarrow \left( 1 - \tilde{x} \right)^2 \partial_{\tilde{x}}$. The Jacobian sparsity pattern for the system of nonlinear equations are first obtained by Maple and the system are then solved numerically by using MATLAB, subject to good intial guesses. The error arises from our numerical results is $\mathcal{O}(10^{-3})$.

\section{Half-Dyon Properties}

In order to define the electromagnetic and neutral $Z_0$ potential, the gauge potentials $A^{a}_{\mu}$ and Higgs field $\Phi^a$ are first gauge transformed to $A^{'a}_{\mu}$ and $\Phi^{a'}$ in the unitary gauge, by using the gauge transformation
\begin{align}
& U = \begin{bmatrix}
     \cos\frac{\alpha}{2} & \sin\frac{\alpha}{2} e^{- i n \phi} \\
     \sin\frac{\alpha}{2} e^{i n\phi}& -\cos\frac{\alpha}{2}
  \end{bmatrix} = \cos\frac{\Theta}{2} + i \hat{u}^{a}_{r} \sigma^a \sin\frac{\Theta}{2}, 
\nonumber\\
& \Theta = - \pi, \nonumber\\ 
& \hat{u}^{a}_{r} = \sin\frac{\alpha}{2} \cos n\phi \, \delta^{a}_{1} + \sin\frac{\alpha}{2} \cos n\phi \, \delta^{a}_{2} +  \cos\frac{\alpha}{2} \delta^{a}_{3}.
\label{eq.32}
\end{align} 
The transformed Higgs column unit vector and the SU(2) gauge potentials in the unitary gauge are
\begin{align}
 \xi' &= U \xi = \begin{bmatrix}
     0 \\
     1  
  \end{bmatrix} \nonumber\\
 g A^{a'}_{\mu} &= - g A^{a}_{\mu} - \partial_{\mu} \alpha \, \hat{u}^{a}_{\phi} - \frac{2 n \sin\frac{\alpha}{2}}{r \sin\theta} \hat{u}^{a}_{\theta} \, \hat{\phi}_{\mu} \nonumber\\
 &- \frac{2n}{r} \left\{ \frac{P_1 \sin\left( \theta - \frac{\alpha}{2} \right)}{\sin\theta} + \frac{P_2 \cos \left( \theta - \frac{\alpha}{2} \right)}{\sin\theta} \right\} \hat{u}^a_{r} \, \hat{\phi}_{\mu}  \nonumber\\
 &+2 \left\{ \tau_1 \cos \left( \theta-\frac{\alpha}{2} \right) - \tau_2 \sin \left( \theta-\frac{\alpha}{2} \right) \right\} \hat{u}^a_{r} \delta^{0}_{\mu},
\label{eq.33}
\end{align} 
which gives
\begin{equation}
g A^{3'}_{\mu} = \left( \tau_1 h_1 + \tau_2 h_2  \right) \delta^{0}_{\mu} + \frac{A_s}{r\sin\theta}  \hat{\phi}_{i} \delta^{i}_{\mu}.
\label{eq.34}
\end{equation} 
Subsequently the electromagnetic potential $A^{\scalebox{.6}{\mbox{em}}}_{\mu}$ and the neutral potential $Z_{\mu}$ are
\begin{align}
\begin{bmatrix}
     {A}^{\scalebox{.5}{\mbox{em}}}_{\mu} \\
    {\cal Z}_{\mu} 
  \end{bmatrix}
&= \frac{1}{\sqrt{g^2+g^{'2}}} \begin{bmatrix}
     g & g' \\
     -g{'} & g 
  \end{bmatrix}  
\begin{bmatrix}
    B_{\mu} \\
    A^{3'}_{\mu}
  \end{bmatrix} \nonumber\\
&= \begin{bmatrix}
     \cos\theta_{\scalebox{.6}{\mbox{W}}} & \sin\theta_{\scalebox{.5}{\mbox{W}}} \\
     -\sin\theta_{\scalebox{.6}{\mbox{W}}} & \cos\theta_{\scalebox{.5}{\mbox{W}}} 
  \end{bmatrix}  
\begin{bmatrix}
    B_{\mu} \\
    A^{3'}_{\mu}
  \end{bmatrix},
\label{eq.35}
\end{align}
where $\cos\theta_{\scalebox{.6}{\mbox{W}}} = g/\sqrt{g^2+g'^2}$ and the electric charge is $e = g g'/\sqrt{g^2 + g'^2}$. Hence the electromagnetic potential and neutral potential can be written as
\begin{align}
A^{\scalebox{.6}{\mbox{em}}}_{\mu} &= \frac{1}{\sqrt{g^2+g'^2}} \left(g' B_{\mu} + g A^{3'}_{\mu}  \right) \nonumber\\
&= \frac{1}{e} \left( \cos^2\theta_{\scalebox{.6}{\mbox{W}}} g' B_{\mu} + \sin^2\theta_{\scalebox{.6}{\mbox{W}}} g A^{'3}_{\mu} \right), 
\label{eq.36}
\end{align} 
and
\begin{align}
Z_{\mu} &=  \frac{1}{\sqrt{g^2+g'^2}} \left( -g' B_{\mu} + g A^{3'}_{\mu}  \right) \nonumber\\ 
&= \frac{1}{e} \sin\theta_{\scalebox{.5}{\mbox{W}}} \cos\theta_{\scalebox{.5}{\mbox{W}}} \left( -g' B_{\mu} +  g A^{'3}_{\mu} \right),
\label{eq.37}
\end{align} 
where
\begin{equation}
g' B_{\mu} = \tilde{B}_0 \delta^{0}_{\mu} + \frac{n B_s}{r \sin\theta} \hat{\phi}_{i} \delta^{i}_{\mu}.
\label{eq.38} 
\end{equation} 

The `em' magnetic field can then be calculated as 
\begin{align}
B^{\scalebox{.6}{\mbox{em}}}_{i} &= -\frac{1}{2} \epsilon_{ijk} F^{\scalebox{.6}{\mbox{em}}}_{jk} \nonumber\\
&= - \frac{1}{e}\epsilon_{ijk} \partial_{j} \left\{ \cos^{2}\theta_{\scalebox{.6}{\mbox{W}}} B_s + \sin^{2}\theta_{\scalebox{.6}{\mbox{W}}} A_s \right\} \partial_{k} \phi,
\label{eq.39}
\end{align} 
and the magnetic field lines can be shown by drawing the lines of constant $\left\{  \cos^{2}\theta_{\scalebox{.5}{\mbox{W}}} B_s + \sin^{2}\theta_{\scalebox{.5}{\mbox{W}}} A_s  \right\} $. The total magnetic charge of the system can be calculated through
\begin{equation}
q_{m} = \int_{S^2} B^{\scalebox{.6}{\mbox{em}}}_{i} r^2 \sin\theta \, d\theta d\phi \, \hat{r}_i ,
\label{eq.40}
\end{equation} 
where $S^2$ is a 2D sphere with radius $r$ centered at the origin. To calculate the magnetic charge enclosed in upper space, we first define a closed surface $S_{+} = H^{2}_{+} \cup D^2$, where $H^{2}_{+}$ is an upper half-sphere with radius $r$ and $D^2$ is a disk with radius $r$ in the $x$-$y$ plane centered at the origin. The magnetic charge enclosed in upper space then is
\begin{equation}
q^{\scalebox{.5}{\mbox{upper}}}_{m} = \int_{S^2_{+}} B^{\scalebox{.6}{\mbox{em}}}_{i} r^2 \sin\theta \, d\theta d\phi \, \hat{r}_i - \int_{D^2} B^{\scalebox{.5}{\mbox{em}}}_{i} r  \, dr d\phi \, \hat{z}_i.
\label{eq.41}   
\end{equation} 
The first term in Eq.(\ref{eq.41}) is evaluated over the upper half-sphere $H^2_{+} $ at $ r \rightarrow \infty$ from $\theta = 0$ to $\theta = \pi/2$, whereas the second term in Eq.(\ref{eq.41}) is evaluated over the horizontal disk $D^2$ at $\theta = \pi/2$ from $r = 0$ to $r = \infty$. Similarly the magnetic charge enclosed in lower hemisphere is
\begin{equation}
q^{\scalebox{.5}{\mbox{lower}}}_{m} = \int_{S^2_{-}} B^{\scalebox{.6}{\mbox{em}}}_{i} r^2 \sin\theta \, d\theta d\phi \, \hat{r}_i + \int_{D^2} B^{\scalebox{.6}{\mbox{em}}}_{i} r  \, dr d\phi \, \hat{z}_i ,
\label{eq.42}
\end{equation} 
where the first term in Eq.(\ref{eq.42}) is evaluated over the lower half-sphere $S^2_{-} $ at $r \rightarrow \infty$ from $\theta = \pi/2$ to $\theta = \pi$.

For the calculation of total electric charge, we consider the time component of gauge potential as 
\begin{equation}
g A^{3'}_{0} = \left( \tau_1 h_1 + \tau_2 h_2  \right), ~g' B_{0} = \tilde{B}_0,
\label{eq.43}
\end{equation} 
which leads to
\begin{equation}
A^{\scalebox{.5}{\mbox{em}}}_{0} = \frac{1}{e} \left[ \cos^2\theta_{\scalebox{.5}{\mbox{W}}} \tilde{B}_0 + \sin^2\theta_{\scalebox{.5}{\mbox{W}}} \left( \tau_1 h_1 + \tau_2 h_2  \right)  \right] .
\label{eq.44}
\end{equation} 
Hence the electric field is
\begin{align}
E^{\scalebox{.5}{\mbox{em}}}_{i} &= F^{\scalebox{.5}{\mbox{em}}}_{i0} = \partial_{i} A^{\scalebox{.5}{\mbox{em}}}_{0} - \partial_{0} A^{\scalebox{.5}{\mbox{em}}}_{i}   \nonumber\\
&= \frac{1}{e} \left[ \cos^2\theta_{\scalebox{.5}{\mbox{W}}} \, \partial_{i}\tilde{B}_0 + \sin^2\theta_{\scalebox{.5}{\mbox{W}}} \,  \partial_{i}\left( \tau_1 h_1 + \tau_2 h_2  \right)  \right] ,
\label{eq.45}
\end{align} 
and the total electric charge of the system is 
\begin{equation}
q_{e} = \int \partial^{i} E^{\scalebox{.6}{\mbox{em}}}_{i} \, dV = \int_{S^2} E^{\scalebox{.6}{\mbox{em}}}_{i} r^2 \sin\theta \, d\theta d\phi \, \hat{r}_i.
\label{eq.46}
\end{equation} 
We use the surface integration in Eq.(\ref{eq.46}) to calculate total electric charge numerically.

The `magnetic' and `electric' neutral field can similarly be written as
\begin{align}
 B^{\scalebox{.5}{\mbox{neutral}}}_{i} &= - \frac{1}{e} \cos\theta_{\scalebox{.5}{\mbox{W}}} \sin\theta_{\scalebox{.5}{\mbox{W}}} \epsilon_{ijk} \partial_{j} \left\{ - n B_s + A_s \right\} \partial_{k} \phi, \nonumber\\
 E^{\scalebox{.5}{\mbox{neutral}}}_{i} &= \frac{1}{e} \cos\theta_{\scalebox{.5}{\mbox{W}}} \sin\theta_{\scalebox{.5}{\mbox{W}}} \partial_{i} \left\{ -\tilde{B}_0 + \tau_1 h_1 + \tau_2 h_2 \right\},
\label{eq.47}
\end{align}
and the `magnetic' neutral field lines can be shown by drawing the lines of constant $\left\{ - n B_s + A_s  \right\} $. The total `magnetic' and `electric' neutral charge of the system are then given by
\begin{align}
 z_{m} &= \int_{S^2} B^{\scalebox{.6}{\mbox{neutral}}}_{i} r^2 \sin\theta \, d\theta d\phi \, \hat{r}_i , \nonumber\\
 z_{e} &= \int_{S^2} E^{\scalebox{.6}{\mbox{neutral}}}_{i} r^2 \sin\theta \, d\theta d\phi \, \hat{r}_i .
\label{eq.48}
\end{align} 

Since the U(1) gauge potential $g' B_{i}$ of the solutions here approaches the 't Hooft gauge potential $g A^{3'}_{i}$ at large $r$, that is $g' B_{i} \approx g A^{3'}_{i}$, the `magnetic' neutral potential $Z_{i}$ therefore vanishes as $r \rightarrow \infty$ and the solution carries zero `magnetic' neutral charge, $z_m \rightarrow 0$. Also the spatial derivative of $Z_0 =  \sin\theta_{\scalebox{.5}{\mbox{W}}} \cos\theta_{\scalebox{.6}{\mbox{W}}} \left( -g' B_{0} +  g A^{'3}_{0} \right) /e $ should approaches zero fast enough asymptotically such that $z_e \rightarrow 0$. Hence asymptotically the solutions should carry net zero $z_m$ and $z_e$. We shall discuss this feature in latter section. 

We can determine the electromagnetic dipole moment of the half-dyon configurations by considering the electromagnetic gauge potential at large $r$,
\begin{equation}
A^{\scalebox{.6}{\mbox{em}}}_{i} \rightarrow \frac{1}{e} \left( g' B_{i} \right) = \frac{1}{e} \frac{n B_s}{r \sin\theta}\hat{\phi}_{i}.
\label{eq.49}
\end{equation} 
Here we perform an asymptotic expansion
\begin{equation}
n B_s \rightarrow - \frac{n}{2} \left( 1 - \cos\theta \right) + \frac{\mu_m \sin^2 \theta}{r},
\label{eq.50}
\end{equation} 
which leads to
\begin{equation}
\mu_m \sin^2\theta = r \left\{n  B_s + \frac{n}{2} \left( 1 - \cos\theta \right) \right\}.
\label{eq.51}
\end{equation} 
By plotting numerical results for the r.h.s of Eq.(\ref{eq.51}), we can then read the value of magnetic dipole moment $\mu_m$ in unit of $1/e$ at $\theta =\pi/2$.

Here we also consider the total angular momentum of the dyonic system. First the energy-momentum tensor of the system is
\begin{align}
T_{\mu\nu} &= F^{\beta a}_{\mu} F^{a}_{\nu \beta} + G^{\beta}_{\mu} G_{\nu \beta}  \nonumber\\
&+ 2 \left( {\cal D}_\mu \phi \right)^{\dagger} \left( {\cal D}_\nu \phi \right) +g_{\mu\nu} {\cal L}.
\label{eq.52}
\end{align} 
From Eq.(\ref{eq.52}) and following Refs. \cite{kn:22,kn:23}, the total angular momentum (along the $z$-direction)
\begin{align}
J_z &= \int T_{0i}\, d^3 x \nonumber\\
&= \int \left\{ F^{a j}_{0} F^{a}_{ij} + G^{j}_{0} G_{ij} + 2 \left( {\cal D}_{0} \phi \right)^{\dagger} \left( {\cal D}_{i} \phi \right) \right\} d^3x
\label{eq.53}
\end{align}
can be reexpressed with help of the equations of motion (\ref{eq.5}) and the symmetry properties of ansatz (\ref{eq.11}) as a surface integral at spatial infinity,
\begin{align}
 J_z = \int_{S^2}& - \left\{ \left( A^{a}_{j} \hat{\phi}_{j} r \sin\theta + \frac{n \delta^{a}_{3}}{g}  \right) F^{a}_{0i} \right. \nonumber\\
&  \left. - \left( B_j \hat{\phi}_{j} r\sin\theta + \frac{n}{g'} \right) G_{0i} \right\}\hat{r}_i  r^2 \sin\theta \, d\theta d\phi.
\label{eq.54}
\end{align}
By considering that as $r \rightarrow \infty$, $\tau_1(r,\theta) \rightarrow \tau(r) \cos (\theta/2)$, $\tau_2(r,\theta) \rightarrow - \tau(r) \sin (\theta/2$ and $\tilde{B}_{0} (r,\theta) \rightarrow B(r)$, the total angular momentum (\ref{eq.54}) becomes
\begin{equation}
J_z = \frac{2 \pi}{g^2} \lim_{r \to \infty} r^2 \partial_r \tau \left( r \right) + \frac{2 \pi}{g'^2} \lim_{r \to \infty} r^2 \partial_r B\left(r \right).
\label{eq.55}
\end{equation}
Now by consider the following expansion at $r \rightarrow \infty$,
\begin{equation}
B = \tau = \eta - \frac{\chi}{r} + O\left( \frac{1}{r^2} \right),
\label{eq.56}
\end{equation}
we obtain 
\begin{equation}
J_z = 2 \pi \left( \frac{\chi}{g^2} + \frac{\chi}{g'^2}  \right) = \frac{2 \pi \chi}{g^2 \sin^2\theta_{\scalebox{.6}{\mbox{W}}}} = \frac{2 \pi}{e^2} \chi.
\label{eq.57}
\end{equation}
For the electric charge, from Eqs.(\ref{eq.45})-(\ref{eq.46}) it can similarly be shown that
\begin{align}
q_e &= \frac{4 \pi}{e} \left[ \cos^2\theta_{\scalebox{.6}{\mbox{W}}} \lim_{r \to \infty} r^2 \partial_r B + \lim_{r \to \infty} r^2 \sin^2\theta_{\scalebox{.6}{\mbox{W}}} \partial_r \tau \right] \nonumber\\
&= \frac{4 \pi}{e} \chi,
\label{eq.58}
\end{align}
as $r \rightarrow \infty$. Hence by comparing Eq.(\ref{eq.57}) and Eq.(\ref{eq.58}), it is obvious that
\begin{equation}
J_z = \frac{q_e}{2 e},
\label{eq.59}
\end{equation}
indicating that the half-dyon solutions possess kinetic energy of rotation.

\section{Results and Discussion}

\begin{figure}[!b]
	\centering
	\hskip0in
	 \includegraphics[width=\linewidth]{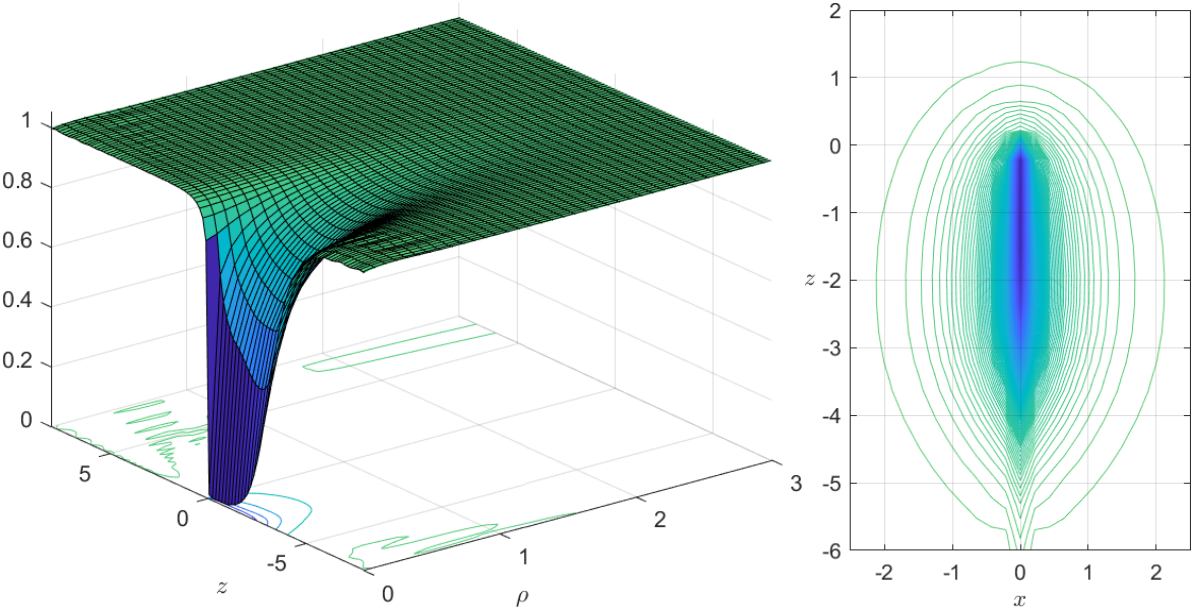} 
	\caption{Higgs modulus and its contour of the Type I half-dyon for $n = 1$ and $\eta = 0.1$. }
\label{higgs-typei}
\end{figure}

By employing correct boundary conditions for Type I half-dyon at $r \rightarrow \infty$, $r = 0$ and $\theta = \pi/2$, analytic calculation of the total magnetic charge from Eqs.(\ref{eq.40})-(\ref{eq.42}) gives
\begin{equation}
q_{m} = \frac{2 n \pi}{e}, ~q^{\scalebox{.5}{\mbox{upper}}}_{m} = 0, ~ q^{\scalebox{.5}{\mbox{lower}}}_{m} = \frac{2 n \pi}{e}.
\end{equation} 
This indicates that the Type I half-dyon configuration possesses magnetic charge of $2 n \pi/e$ that resides only in the lower-space. The $\phi$-winding number $n$ runs only from zero to three, as there are no solutions for $n > 3$. The half-dyon solutions also possess positive electric charge $q_{e}$ that depends on $\eta$. In particular the solutions exist from $\eta = 0$ up to a critical value $\eta_c$, and cease to exist when $\eta > \eta_c$. The critical values for different cases of $n$ are given by $\eta_{c,n=1} = 0.5001, \eta_{c,n=2} = 0.5005$ and $\eta_{c,n=3} = 0.1726$. The net neutral charge $z_m$  and $z_e$ of the system are zero as expected.

\begin{figure}[!tb]
	\centering
	\hskip0in
	 \includegraphics[width=\linewidth]{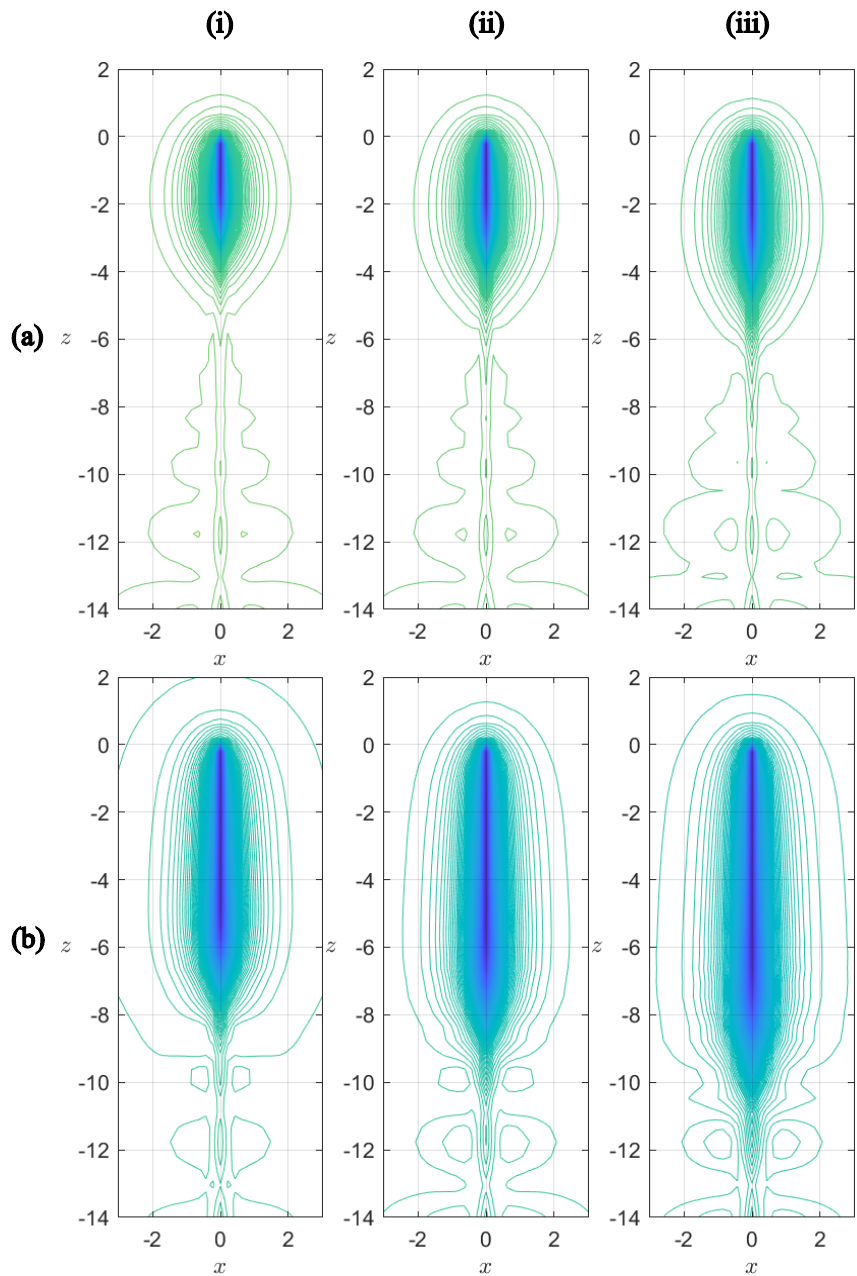} 
	\caption{Higgs modulus contour of the Type I half-dyon for (a) $n = 1$ and (b) $n = 2$, at (i) $\eta = 0$, (ii) $\eta = 0.3$, and (iii) $\eta = 0.4$. }
\label{Fig.3}
\end{figure}

The Higgs modulus and its contour of the Type I half-dyon solution when $n = 1, \eta = 0.1$ are plotted in Fig.~\ref{higgs-typei}. Manifests as an inverted cone, the Higgs modulus shows that the half-dyon resides in lower-space as a finite length object along the negative $z$-axis. From Fig. \ref{Fig.3}, the Higgs modulus contour becomes more elongated (or more blown) along the negative $z$-axis, indicating the increase of the half-dyon's length. Hence, the shape of the half-dyon generally remains the same as that of the half-monopole in Weinberg-Salam model \cite{kn:25}. The increase in size of the half-dyon along the negative $z$-axis as $\eta$ increases also shares similar picture as that of half-dyon in SU(2) YMH theory \cite{su2-half}.

\begin{figure}[!tb]
	\centering
	\hskip0in
	 \includegraphics[width=\linewidth]{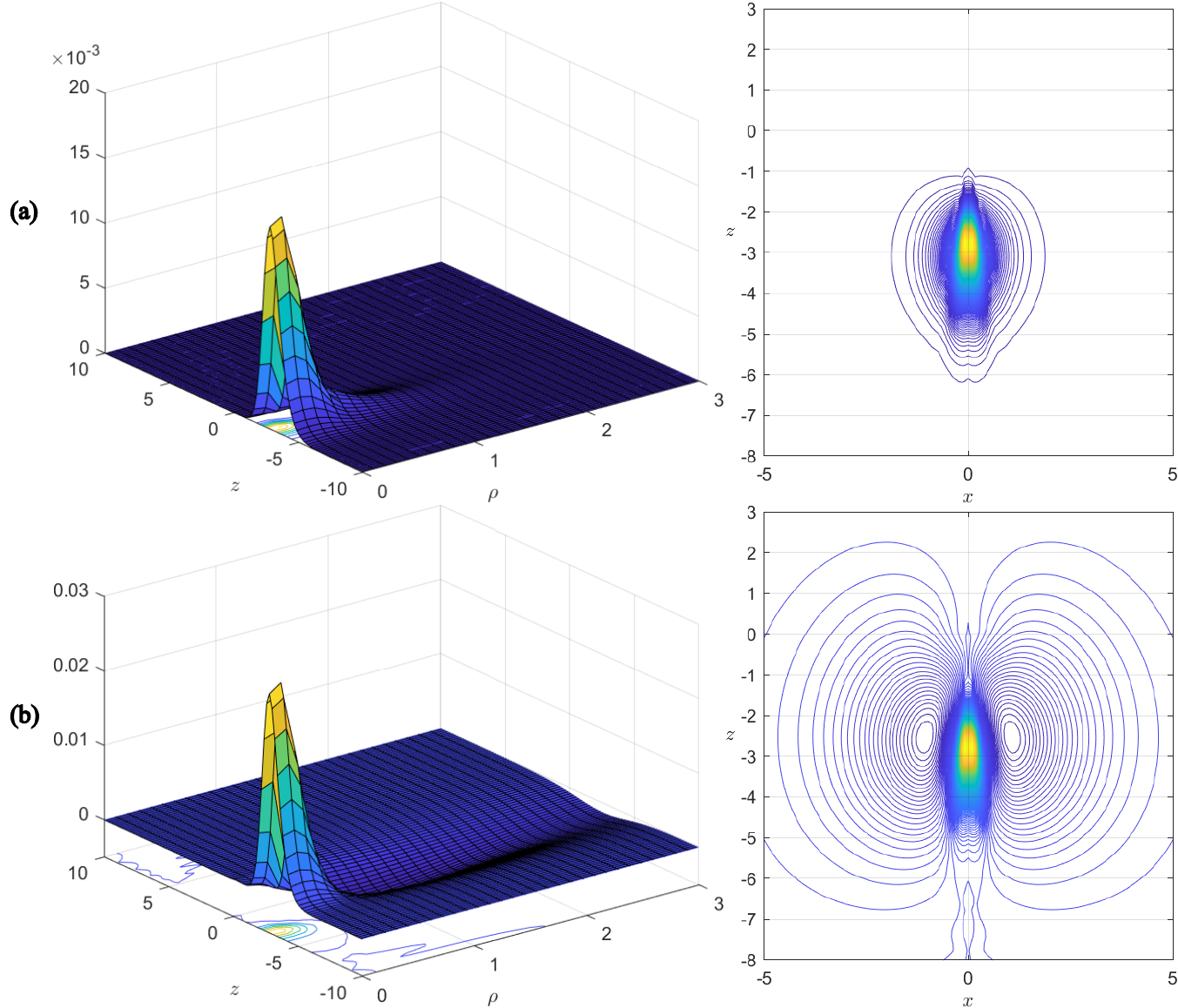} 
	\caption{(a) Weighted electric charge density with its contour; and (b) weighted `electric' neutral charge density with its contour, of the half-dyon solutions with $n = 1, \eta = 0.1$. }
\label{Fig.4}
\end{figure}

\begin{figure}[!t]
	\centering
	\hskip0in
	 \includegraphics[width=\linewidth]{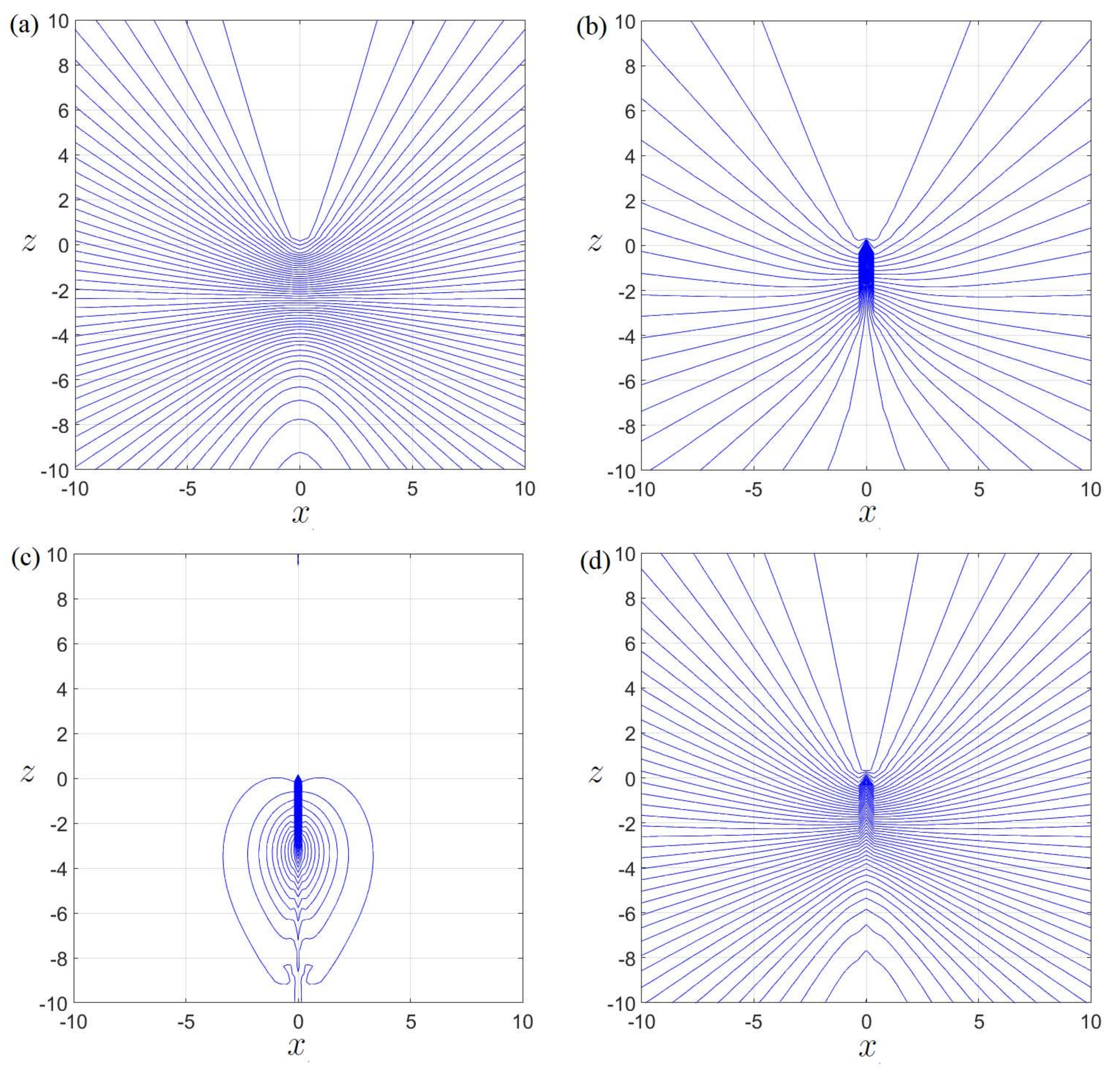} 
	\caption{(a) U(1) magnetic field lines, (b) SU(2) magnetic field lines, (c) `magnetic' neutral field lines and (d) `em' magnetic field lines of the half-dyon solutions with $n = 1, \eta = 0.1$ .}
\label{Fig.5}
\end{figure}

From Fig. \ref{Fig.4}, it is obvious that besides magnetic charge density, the half-dyon carries electric charge density that are concentrated along the negative $z$-axis near the origin, together with `electric' neutral charge density. The magnetic field lines and `magnetic' neutral field lines of the half-dyon are shown in Fig. \ref{Fig.5}. While the magnetic field lines represent that of an object with finite magnetic charge $q_m$ along the negative $z$-axis extending from the origin, the plot of `magnetic' neutral field lines suggests that there are finite $z_m$ concentration in the lower space. The upper part possesses positive $z_m$ whereas the lower part possesses negative $z_m$ adjacently, giving an overall net zero `magnetic' neutral charge for the system. 

\begin{figure}[!tb]
	\centering
	\hskip0in
	 \includegraphics[height=0.9\textheight]{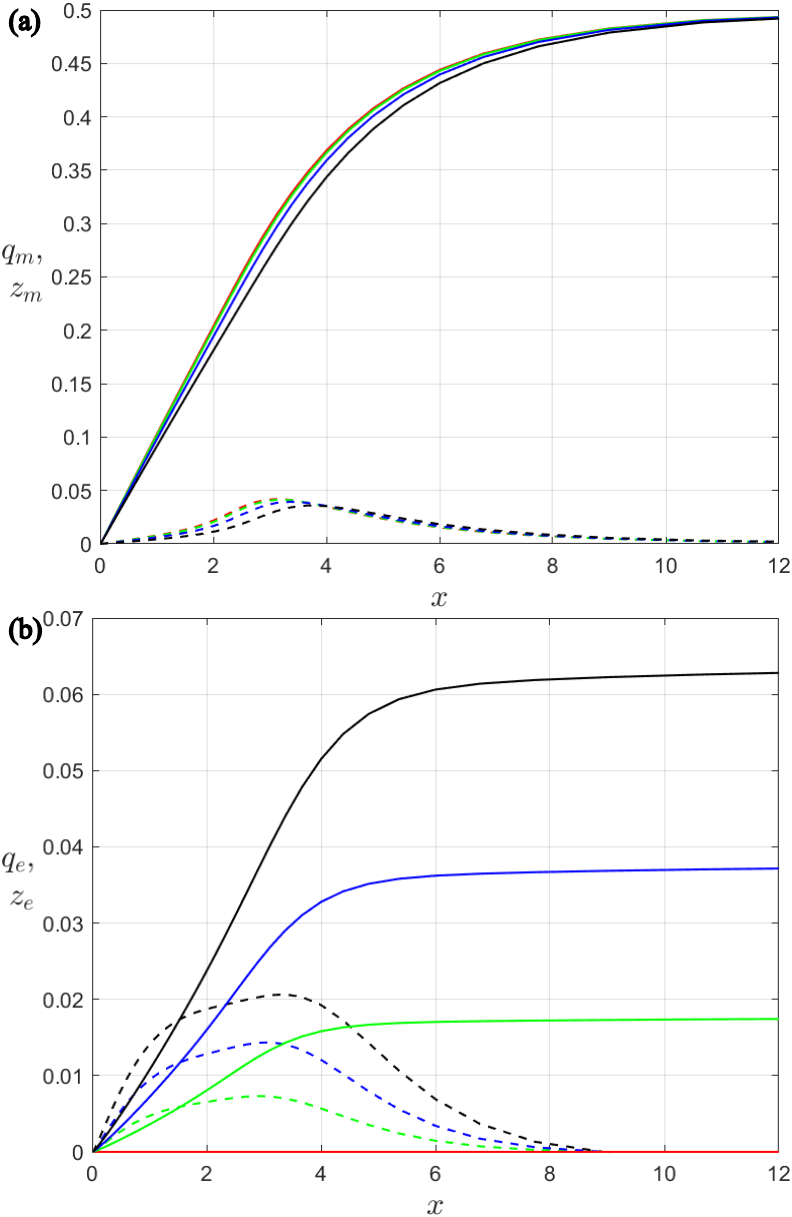} 
	\caption{Functions of (a) magnetic charge $q_m \left( x \right)$ (solid) and `magnetic' neutral charge $z_m \left( x \right)$ (dash-dotted), and (b) electric charge $q_e \left( x \right)$ (solid) and `electric' neutral charge $z_e \left( x \right)$ (dash-dotted) versus $x$; for the Type I half-dyon solutions with $n = 1$,  $\eta = 0$ (red), $\eta = 0.2$ (green), $\eta = 0.3$ (blue), and $\eta = 0.4$ (black).  }
\label{Fig.6}
\end{figure}

To confirm the exact picture of the charges, we consider again Eqs.(\ref{eq.40}), (\ref{eq.46}) and (\ref{eq.48}), with the radius of $S^2$ now varies with dimensionless coordinate $x$. The functions of $q_m(x), q_e(x), z_m (x)$ and $z_e (x)$ in units of $4 \pi/e$ versus $x$ of half-dyon with $n =1$ and various $\eta$ are then plotted in Fig. \ref{Fig.6}. As expected, $q_m$ and $q_e$ rises from zero at $x = 0$ to $2 \pi/e$ and maximal $q_{e}$ respectively as $x \rightarrow \infty$. For $z_m$ and $z_e$, they both rises from zero at $x =0$ to a maximal value at intermediate $x$ before declining to zero as $x \rightarrow \infty$. This shows that there exist equal amount of positive and negative charges at intermediate $x$ but as $x \rightarrow \infty$, $z_m \rightarrow 0$ and $z_e \rightarrow 0$. These phenomenons are also observed for the case of $n=2, 3$ and similar statement has also been presented in Section 4.

\begin{figure}[!tb]
	\centering
	\hskip0in
	 \includegraphics[width=\linewidth]{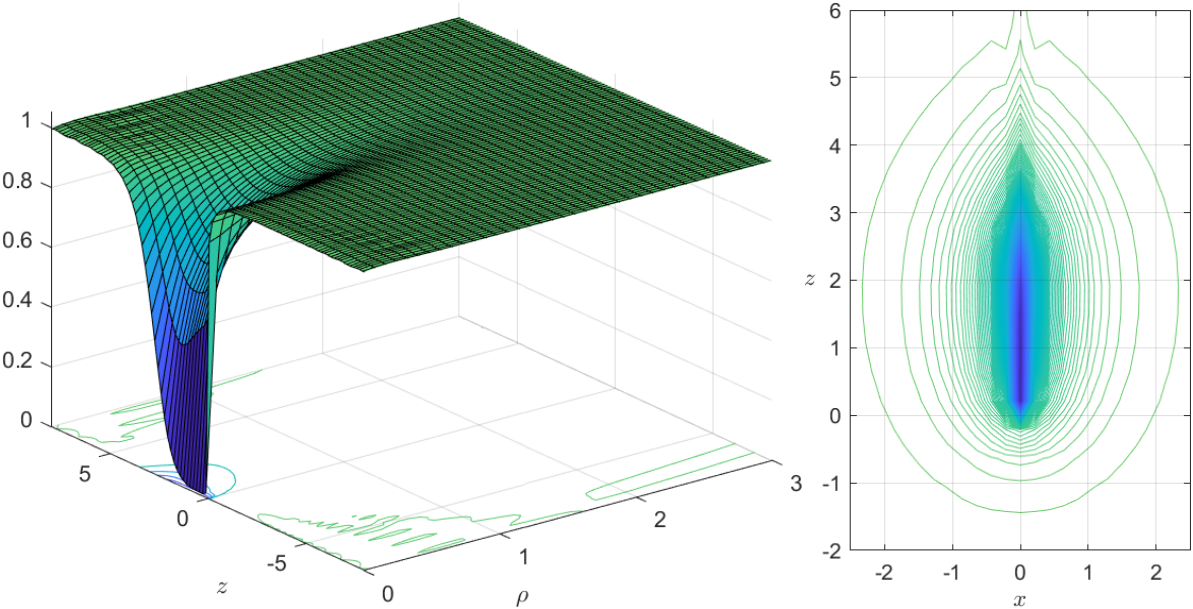} 
	\caption{Higgs modulus and its contour of the Type II half-dyon for $n = 1$ and $\eta = 0.1$. }
\label{higgs-typeii}
\end{figure}

\begin{figure}[!tb]
	\centering
	\hskip0in
	 \includegraphics[width=\linewidth]{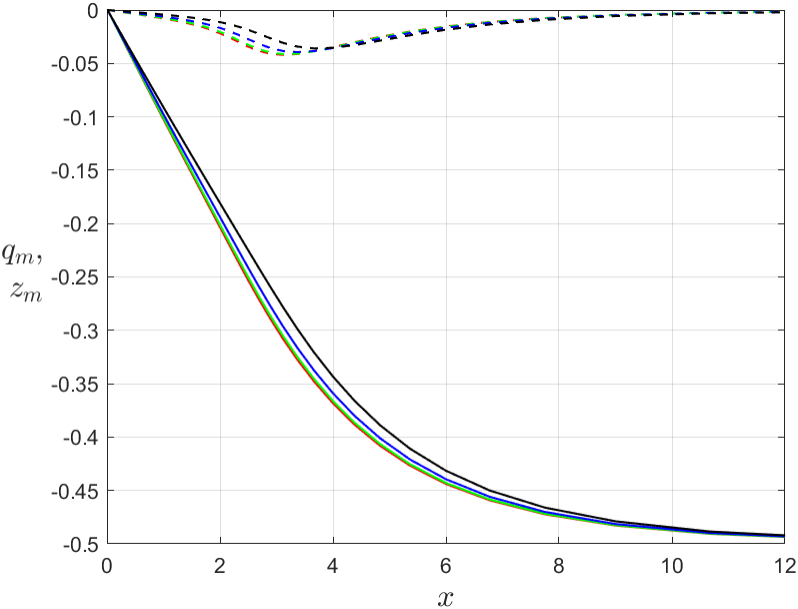} 
	\caption{Functions of magnetic charge $q_m \left( x \right)$ (solid) and `magnetic' neutral charge $z_m \left( x \right)$ (dash-dotted) for the Type II half-dyon solutions with $n = 1$,  $\eta = 0$ (red), $\eta = 0.2$ (green), $\eta = 0.3$ (blue), and $\eta = 0.4$ (black).  }
\label{magnetic-typeii}
\end{figure}

\begin{figure*}[!b]
	\centering
	\hskip0in
	 \includegraphics[width=\textwidth,keepaspectratio]{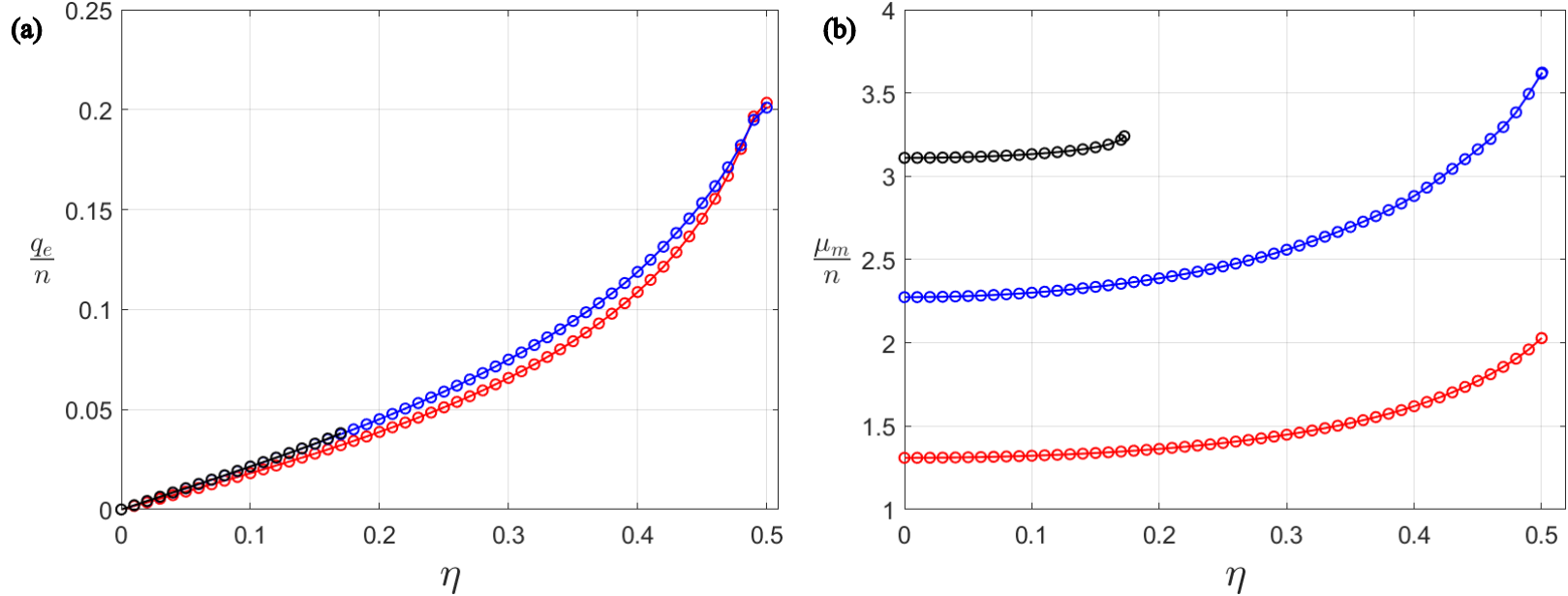} 
	\caption{Plots of (a) $q_e / n$ (in unit of $4 \pi /e$) versus $\eta$, (b) $ \mu_m/n$ (in unit of $1 /e$) versus $\eta$; of the Type I half-dyon solutions for $n = 1$ (red),  $n=2$ (blue) and $n = 3$ (black).   }
\label{Electric-Moment-plot}
\end{figure*}

From the above statement, this means that the conditions of $z_m$ and $z_e$ approaching zero at large $x$ do not indicate that the neutral charge densities are zero over all space, but they exist in intermediate region. Again from Fig. \ref{Fig.6}, the strength of these neutral charge densities in intermediate region are noticeably small, but not negligible. One may wonder if these are numerical errors but repeated calculations show that they are not. A possible remedy is to increase the accuracy of numerical solutions, i.e. consider a grid size of, say $120 \times 110$. This will be carried out in another work, but that will mean consumption of gigantic computational resources. In short the occurrence of neutral charge densities in the half-dyon solutions would demand further in-depth investigations.

Hence our solutions indicate that if a monopole undergoes distortion (due to excitation), neutral charge densities (with minute strength) would emerge, but the overall neutral charges would remain zero. There is however a crucial difference between $z_m$ and $z_e$. For $z_m$, its positive charge and negative charge are adjacently distributed along the negative $z$-axis, Fig. \ref{Fig.5}(c). For $z_e$, the positive charges are distributed along the negative $z$-axis extending from the origin, while its negative charge are distributed along a finite tube encircling the positive charge, Fig. \ref{Fig.4}(b).

\begin{table}[!b]

\begin{adjustbox}{width=1\textwidth}
\label{tab:1}       % Give a unique label
% For LaTeX tables use
\begin{tabular}{llllllllllll}
\hline\noalign{\smallskip}
$n=1$ &  &  &  &  &  &  & & & & &\\
\hline\noalign{\smallskip}
$\eta$ & 0 & 0.05 & 0.10 & 0.15 & 0.20 & 0.25 & 0.30 & 0.35 & 0.40 & 0.45 & 0.5001\\
\noalign{\smallskip}\hline\noalign{\smallskip}
$q_e $ & 0 & 0.0089 & 0.0181 & 0.0280 & 0.0388 & 0.0511 & 0.0658 & 0.0842 & 0.1088 & 0.1455 & 0.2167 \\ 
$ \mu_m/n $ & 1.3101 & 1.3131 & 1.3225 & 1.3389 & 1.3637 & 1.3991 & 1.4488 & 1.5187 & 1.6196 & 1.7718 & 2.0291 \\ 
\hline\noalign{\smallskip}

\hline\noalign{\smallskip}
$n=2$ &  &  &  &  &  & & & & & &\\
\noalign{\smallskip}\hline\noalign{\smallskip}
$\eta$ & 0 & 0.05 & 0.10 & 0.15 & 0.20 & 0.25 & 0.30 & 0.35 & 0.40 & 0.45 & 0.5005\\
\noalign{\smallskip}\hline\noalign{\smallskip}
$q_e $ & 0 & 0.0203 & 0.0412 & 0.0632 & 0.0871 & 0.1138 & 0.1449 & 0.1825 & 0.2306 & 0.2989 & 0.4195 \\ 
$\mu_m / n$ & 2.2744 & 2.2811 & 2.3014 & 2.3364 & 2.3878 & 2.4589 & 2.5594 & 2.6960 & 2.8816 & 3.1617 & 3.6229\\ 
\hline\noalign{\smallskip}

\hline\noalign{\smallskip}
$n=3$ &  &  &  &  &  & & & & & &\\
\noalign{\smallskip}\hline\noalign{\smallskip}
$\eta$ & 0 & 0.05 & 0.10 & 0.15 & 0.1726 & & & & & & \\
\noalign{\smallskip}\hline\noalign{\smallskip}
$q_e $ & 0 & 0.0310 & 0.0627 & 0.0961 & 0.1141 &  &  &  & & &\\ 
$\mu_m /n $ & 3.1107 & 3.1157 & 3.1329 & 3.1739 & 3.2407 & &  &  & & &\\ 
%$\tilde{E}$ & 3.6969 & 3.7006 & 3.7117 & 3.7309 & 3.7443 &  &  &  & & &\\ 
%$\tilde{E}/n$ & 1.2323 & 1.2335 & 1.2372 & 1.2436 & 1.2481 &  &  &  & & &\\ 
%$E$ & 18.1560 & 18.1741 & 18.2287 & 18.3230 & 18.3888 &  &  & & & &\\ 
\noalign{\smallskip}\hline

\end{tabular}
\end{adjustbox}
\caption{Selected values for electric charge $q_e$ in unit of $4\pi/e$ and magnetic dipole moment per $n$ ($\mu_m/n$) in unit of $1/e$ of the Type I half-dyon solution at physical Weinberg angle $\sin^2\theta_{\scalebox{.5}{\mbox{W}}} = 0.2229$ and Higgs self-coupling constant $\beta = 0.77818833$ for $\phi$-winding number $1 \leq n \leq 3$.}
\end{table}

Next, the Type II solution is obtained by applying the boundary conditions (\ref{eq.26}), (\ref{boundary-cond-negative_1}), (\ref{boundary-cond-negative_2}), and (\ref{boundary-cond-negative_3}). We similarly perform analytic calculation of total magnetic charge of Type II solutions using Eqs.(\ref{eq.40})-(\ref{eq.42}) and obtain 
\begin{equation}
q_{m} = -\frac{2 n \pi}{e}, ~q^{\scalebox{.5}{\mbox{upper}}}_{m} = -\frac{2 n \pi}{e}, ~ q^{\scalebox{.5}{\mbox{lower}}}_{m} = 0.
\end{equation} 
This implies that the Type II half-dyon solution contains magnetic charge of $-2n\pi/e$ and distributes along the positive $z$-axis. Type II solution clearly differs from the Type I solution which has $+2n\pi/e$ and lies along the negative $z$-axis. Fig.~\ref{higgs-typeii} shows the Higgs modulus and its contour of Type II half-dyon solution at $n=1$, $\eta=0.1$.   Both the magnetic and `magnetic' neutral charges of Type II half-dyons are plotted in Fig.~\ref{magnetic-typeii} which confirms our analysis that the magnetic charges are negative-valued. 

The Type II solutions have the exact same critical value $\eta_c$ as their corresponding Type I solutions, which are $0.5001$, $0.5005$ and $0.1726$ for $n = 1, 2$ and 3 respectively. Despite the opposite magnetic charge, both types of solutions share identical values for electric $q_e$ and `electric' neutral $z_e$ charges. Comparison of Fig.~\ref{higgs-typei} and Fig.~\ref{higgs-typeii} demonstrates that, in terms of structure, Type II half-dyons are perfect mirror images of Type I half-dyons across the $\rho=\sqrt{x^2+y^2}$ plane. These findings imply that the emergence of magnetic monopoles (or dyons) may occur in pairs, hinting at the potential for monopole pair production.

The plots of total electric charge per $n$, $q_e/n$ and magnetic dipole moment per $n$ ($\mu_m / n$) versus $\eta$ are shown in Fig. \ref{Electric-Moment-plot}. In general, the solutions possess a minimum value for $q_e/n$ and $\mu_m/n$ when $\eta = 0$, followed by an (approximately) quadratic increase until they reach their critical values at $\eta = \eta_{c}$. Hence overall the half-dyon solutions in Weinberg-Salam model possess almost similar pattern as the half-dyon in SU(2) YMH theory \cite{su2-half}. This is not surprising since the Weinberg-Salam theory contains the SU(2) group. Selected values of $q_e$ and $\mu_m/n$ for the Type I half-dyon configurations are tabulated in Table 1 for $n = 1$ to $n = 3$. For Type II half-dyon, it has exactly the same electric charge, `electric' netural charge, and magnetic dipole moment as Type I half-dyon. Hence giving the same graphical behavior.

Here we also elaborate on the Type I half-dyon solution with $\eta = 0, n = 2$, which is basically an electrically neutral monopole system that possesses magnetic charge $4\pi/e$. In comparison with the spherically symmetric Cho-Maison monopole \cite{kn:8} (which also possesses magnetic charge $4\pi/e$), our solution with axial symmetry can be viewed as a superposition of two half-monopoles (hereon denoted as two-half-monopole). Besides symmetry, there exists some differences between the two-half-monopole and the Cho-Maison monopole. First the Cho-Maison monopole has absolutely no neutral charge (`electric' and `magnetic'). The two-half-monopole here possesses net zero neutral charge, as dictated by the boundary condition (\ref{eq.25}). However at finite $r$ there are neutral charge densities which consists of equal amount of positive and negative neutral charge. This indicates the distortion of Cho-Maison monopole might lead to emergence of neutral charge density, though asymptotically the total neutral charge is zero.

\section{Conclusions}
\label{sec:6}

We have studied numerical solutions in Weinberg-Salam theory corresponding to axially symmetric electrically charged half-monopole system. Here we considered the conditions of physical Weinberg angle $\sin^2\theta_{\scalebox{.6}{\mbox{W}}} = 0.2229$, Higgs self-coupling constant $\beta = 0.77818833$ and $\phi$-winding number $1 \leq n \leq 3$. Our analysis, considering the correct dimensions of the theory, thoroughly explores fundamental properties such as magnetic charge, neutral charge, and magnetic dipole moment for both half-dyon solutions. These generalized half-dyon in Weinberg-Salam theory are quite similar to the generalized half-dyon in SU(2) Yang-Mills-Higgs theory~\cite{su2-half}, where they will also be categorized as Type I and Type II solutions in this paper.

The Type I half-dyon solution in a finite length object distributed along the negative $z$-axis extending from the origin. They possess net magnetic charge $2 n \pi/e$, electric charge ranging from zero up to a maximum $q_e$ and zero net neutral (electric and magnetic) charge. Conversely, the Type II half-dyon is the same object as Type I half-dyon but distributed along the positive $z$-axis. The Type II half-dyon solutions carry a net magnetic charge of $-2n\pi/e$, mirroring the maximum electric charge $q_e$ and zero net neutral (electric and magnetic) charge of their Type I counterparts under identical conditions. Despite approaching zero at large distances, the half-dyon solutions reveal the presence of neutral (electric and magnetic) charge densities at intermediate regions, albeit with notably small strengths, prompting a need for further extensive investigations. The presence of both Type I and Type solutions provides support for the idea that magnetic monopoles (or dyons) could appear in pairs, thereby signifying the possibility of magnetic monopole pair production.

The total energy of the half-dyon system is infinite and cannot be calculated due to the singularity arises from the U(1) part of energy density. However it is possible to regularize the solution by introducing a non-trivial U(1) hypercharge permeability in the Lagrangian \cite{kn:8,kn:9,kn:10}. Further study on this regularization method will be reported elsewhere. Solutions in SU(2) $\times$ U(1) Weinberg-Salam theory can also be coupled to gravity \cite{kn:27,kn:28}. Further work to study the half-monopole (half-dyon) system under the influence of gravity is under way and will be reported in another work.

\end{document}